\documentstyle[proceedings]{crckapb} 

\begin{opening}
\title{COUNTERROTATION IN GALAXIES}
\author{F. BERTOLA}
\author{E.M. CORSINI}
\institute{Dipartimento di Astronomia, Universit\`a di Padova\\
           vicolo dell'Osservatorio 5, I-35122 Padova, Italy}
\end{opening}

\begin{document}

\section{Introduction} 

The phenomenon of counterrotation is observed when two galaxy components
have their angular momenta projected antiparallel onto the sky. It follows
that if the two components rotate around the same axis, the counterrotation
is intrinsic. On the contrary the counterrotation is only apparent if the
rotation axes are misaligned and the line-of-sight lies in between the two
vectors or their antivectors. In the case of intrinsic counterrotation the
two components can be superimposed or radially separated. 

As far as the two components are concerned, stars are observed to
counterrotate with respect to other stars or gas. The counterrotation of
gas versus gas has been also detected. Up to now,  the number of galaxies
exhibiting these phenomena are $\sim60$, the morphological type of which
ranges from ellipticals to S0's and to spirals. Previous reviews about
counterrotation are those of Rubin (1994b) and Galletta (1996). 

When a second event occurs in a galaxy, such as the acquisition of material
from outside, it is likely that the resulting angular momentum of the
acquired material is decoupled from the angular momentum of the preexisting
galaxy. Counterrotation is therefore a general signature of material
acquired from outside the main confines of the galaxy. Good examples of
such cases are ellipticals with a dust lane or gaseous disk along the minor
axis and polar ring galaxies, where the angular momenta are perpendicular.
It should be noted that recently attempts have been made to explain special
cases of stars versus stars counterrotation in disk galaxies as due to a
self induced phenomenon in non-axisymmetric potentials (Evans \& Collett
1994, Wozniak \& Pfenninger 1997). 

In the following we discuss the phenomenon according to the morphological
type and to the kind of counterrotation.

\section{Elliptical Galaxies}

\subsection{Gas vs. Stars}

Disks of ionized gas, which appear as dust lanes crossing the stellar body
when seen on edge, have been detected in a large fraction ($\sim50$\%) of
ellipticals (e.g. Macchetto et al. 1996). Typical masses of dust and
ionized gas in ellipticals are M$_{\rm dust}=10^{5}-10^{6}$ M$_{\odot}$ and
M$_{\rm HII}=10^{3}-10^{5}$ M$_{\odot}$ (e.g. Bregman et al. 1992). The
kinematical decoupling generally observed (Bertola et al. 1990) between the
gaseous and the stellar components suggests the gas is settled or in the
process of settling in the equilibrium configurations. In triaxial
ellipticals they are the planes orthogonal to the shortest and to the
longest axes. 

Intrinsic counterrotation of gas and stars in ellipticals has been observed
in Anon~1029-459 (Bertola et al. 1988), IC~2006 (Schweizer et al. 1989),
NGC~3528 (Bertola et al. 1988), NGC~5354 (Bettoni et al. 1995), NGC~5898
(Bettoni 1984, Bertola \& Bettoni 1988), and NGC~7097 (Caldwell et al.
1986).

\subsection{Stars vs. Stars}

Kinematically decoupled cores are observed in ellipticals when their
stellar velocity field show a discontinuity between the rotation of the
inner and the outer regions. Isolated cores have been detected in 17
ellipticals (Mehlert et al. 1997 and references therein) and in IC~4889
(Bertola et al. 1992). About half of them are stellar nuclear disks
(Carollo et al. 1997). 

Stellar counterrotation characterizes IC~1459 (Franx \& Illingworth 1988),
IC~4889 (Bertola et al. 1992), NGC~1439, NGC~1700 (Franx et al. 1989),
NGC~3608 (Jedrzejewsky \&
Schecther 1988), NGC~4472 (Davies \& Birkinshaw 1988), 
NGC~5322 (Bender 1988),
and NGC~4816 (Mehlert et al. 1997), 
NGC~7796 (Bertin et al. 1996). 
At least in the first seven galaxies the
counterrotation is intrinsic as indicate by the lack of velocity gradient
along the minor axis. Minor axis observations are not available for the
remaining two galaxies. 

Several formation scenario have been proposed. They are the dissipationless
minor merging with a compact low-luminosity elliptical (Kormendy 1984,
Balcells \& Quinn 1990), the accretion of gas-rich dwarf companions and
subsequent star formation as in dust-lane ellipticals (Franx \& Illingworth
1988), the dissipational major merging between an elliptical and a disk
galaxy or between two disk galaxies (Schweizer 1990, Hernquist \& Barnes
1991), the hierarchical merging of dynamically hot, but still partially
gaseous objects (Bender \& Surma 1992), and the interaction involving an
elliptical with an embedded disk (Hau \& Thomson 1994).

\subsection{Gas vs. Gas}

In the E4 galaxy NGC~1052 two dimensional high resolution H$\alpha$
spectroscopy has revealed in the central $\pm2$ kpc two apparently
counterrotating ionized gas components (Plana \& Boulesteix 1996). The two
components are interpreted as produced by two distinct structures
superimposed by a projection effect and settled onto the two orthogonal
equilibrium planes of a triaxial elliptical galaxy. They have a mass of
M$_{\rm HII, 1}\sim10^{3}$ M$_{\odot}$ and M$_{\rm HII, 2}\sim10^{5}$
M$_{\odot}$ respectively. 

The case of the E5 galaxy IC~4889 (Vega et al. 1997, these proceedings) is
more complex. In addition to the above mentioned counterrotating core, two
gas disks settled onto the allowed planes of a triaxial galaxy apparently
counterrotate. Both cases indicate multiple acquisition events.

\section{Disk Galaxies}

\subsection{Gas vs. Stars}

A compilation of 36 S0's galaxies with ionized gas based on the samples of
Bertola et al. (1992) and of Kuijken et al. (1996) lists 12 objects with
gas counterrotating with respect to the stars. 
The large fraction of S0's with kinematical decoupling between stars and
gas is consistent with the idea that the gas in lenticulars is acquired
from outside (Bertola et al. 1992). Episodic infall of external gas,
continuous infall of external gas and dissipational merging with a gas-rich
dwarf companion all have been investigated by means of numerical simulation
(Thakar \& Ryden 1996, Thakar et al. 1997) as viable mechanisms for
producing an overall gaseous counterrotation in disk galaxies. In the
infall case the acquisition rate, the initial angular momentum and the gas
clumpiness state are crucial parameters to successfully build up the
counterrotating gas disk without dynamically heating the preexisting
stellar disk. The same is true for the mass of the captured satellite in
the merging case. 

In the edge-on peanut S0 NGC~128 (Emsellem \& Arsenault 1997) a tilted gaseous
disk counterrotates with respect to the stars. It has been interpreted as due
to acquired gas settled onto the so-called anomalous orbits in a tumbling
triaxial potential, as in the case of the barred S0's NGC~2217 (Bettoni et al.
1990) and NGC~4684 (Bettoni et al. 1993). 

In the early Sa NGC~3626 (Ciri et al. 1995) the neutral, ionized and
molecular gas (Garcia-Burillo et al. 1997), amounting to $\sim 10^9$
M$_{\odot}$, counterrotates at all radii with respect to the stars. It
should be noted that, in spite of the presence of two faint amorphous dust
and possibly stellar spiral arms, the morphology of NGC~3626 resembles that
of more typical S0 galaxies.

\subsection{Stars vs. Stars}

\subsubsection{Stellar Disk vs. Stellar Disk}

The S0 galaxy NGC~4550 has two cospatial counterrotating stellar disks, one
of them corotating with the gaseous one (Rubin et al. 1992). The two disks
have exponential surface brightness profiles with the same central surface
brightnesses and scale lengths (Rix et al. 1992). In the early-type spirals
NGC~4138 (Jore et al. 1996) and NGC~7217 (Merrifield \& Kuijken 1994) about
20\% -- 30\% of the stars in the disk are in retrograde orbits constituting
a kinematically distinct component. Both disks are equally extended but
have different levels of surface brightness. In NGC~7217 the gaseous disk
rotates in the same direction as the primary (i.e. the more massive)
stellar disk. The contrary is true for NGC~4138. 

The presence of a counterrotating stellar disk is interpreted as the result
of a subsequent stellar formation in an accreted gaseous disk. In the case
of NGC~7217 where the ionized gas rotates in the same direction as the more
massive disk, Merrifield and Kuijken (1994) suggested that the actual
retrograde (i.e. the less massive) disk formed first.\\ 
An alternative to the second event scenario has been proposed for NGC~4550
by Evans \& Collett (1994). They stated that whenever oval disks become
circular or triaxial halos in which they are embedded become axysimmetric,
then box orbits are scattered equally into clockwise and counterclockwise
rotating tube orbits. In this way two identical counterrotating stellar
disks can be built. The presence of three gaseous rings in NGC~7217 has
suggested to Athanassoula (1996), that in this galaxy the bar has decayed,
causing the counterrotation of the less massive stellar disk. 

Extended stellar counterrotation in disks of S0's and spirals seems to be a
rare phenomenon according to the results of Kuijken et al. (1996). They
measured carefully the line-of-sight velocity distribution along the major
axes of 28 S0 galaxies, without finding any new case. Indeed they estimated
that less than 10\% of S0's show large-scale counterrotation with more than
5\% of disk stars on retrograde orbits. If counterrotating stars formed
from acquired gas, this result contrasts with the large fraction of S0's
exhibiting counterrotating disks of ionized gas. However, most of these gas
disks are too small to produce the large-scale counterrotation observed in
systems like NGC~4138, NGC~4550 and NGC~7217. 

The Sa galaxy NGC~3593 (Bertola et al. 1996) is composed by a small bulge,
a first, radially more concentrated, stellar disk that contains about 20\%
of the total luminous mass and dominates the star kinematics in the inner
$\pm1$ kpc, and a second counterrotating stellar disk, radially more
extended, dominating the outer kinematics. The two disks have exponential
luminosity profiles with different scale lengths and central surface
brightnesses. The gaseous disk counterrotates with respect to the higher
scale length disk. 
The existence of cases like NGC~3593, where two disks of different scale
lengths and central surface brightnesses counterrotate suggest that they
can be numerous if not a general case. The Fig.~1 show the vast area to be
explored in the plane central surface brightness vs. scale length within
the limit of present day detectability. 

\begin{figure}
\vspace{5cm}
\caption{
Detectability of two counterrotating stellar disks having exponential
luminosity profiles of scale lengths $r_1$ and $r_2$ and central
intensities $I_1$ and $I_2$. We assume to disentangle the two disks if the
intensity of the smaller one (disk 2) is at least the 5\% of that of the
greater one (disk 1) at all radii lower than $r_1$ (non-hatched area). The
dots correspond to the observed cases of NGC~4550 (Rubin et al. 1992, Rix
et al. 1992) and NGC~3593 (Bertola et al. 1996).} 
\end{figure}

Instabilities in disk galaxies with counterrotating stars and/or gas have
been recently investigated by Sellwood \& Merritt (1994) and Lovelace et
al. (1997). Due to the results to their numerical simulations Comins et al.
(1997) suggest that one-armed spiral features may characterize disk
galaxies with counterrotation. Spirals with tightly-coiled narrow arms are
the candidates for Howard et al. (1997).

\subsubsection{Stellar Bulge vs. Stellar Disk}

In the Sb NGC~7331 (Prada et al. 1996) the bulge counterrotates with
respect to the disk. The galaxy appears morphologically smooth and
undisturbed, but the analysis of the stellar LOSVD reveals the presence in
the inner $\pm1.4$ kpc of two counterrotating stellar components.
Near-infrared photometry shows that the radial surface brightness profile
of the slow-rotating component follows that of the bulge, while the
fast-rotating follows the disk. Also the Sb spiral NGC~2841 seems to have a
counterrotating bulge (Prada 1997, private communication). 

Did disk galaxies really start out as `undressed spheroids' and the disks
formed gradually over several several billion of years as suggest by Binney
\& May (1986)? To answer this question we need to know how unique are the
cases of NGC~2841 and NGC~7331. The absence of a velocity gradient in the
central ($\pm1$ kpc) regions of the Sa NGC~4698 would suggest the presence
of a counter or orthogonally rotating bulge (Bertola et al. 1997, in
preparation).

\subsubsection{Stellar Counterrotation in Early-Type Barred Galaxies}

The stellar rotation curve along the major axis of the bar of eleven
early-type barred galaxies seen at intermediate inclination show a `waving
pattern'. Nine are listed by Bettoni \& Galletta (1997) while two, namely
NGC~5005 and NGC~5728 were found by Prada (1997, private communication).
The observed amplitude of the oscillations is lower than 30 $\rm
km\;s^{-1}$ producing in the rotation curve a region of minimum, which
sometimes can reach negative values. 

The phenomenon, which was pointed out by Bettoni (1989), has been recently
interpreted by Wozniak \& Pfenninger (1997). They used
self-con\-si\-ste\-nt models of barred galaxies with reproduce the observed
wavy pattern as due to the presence of retrograde orbits with a local
concentration in the region of minimum velocity. 
The origin of such a stellar counterrotation is not necessarily external
but acquired gas can be trapped on this family of  retrograde orbits and
then eventually form stars. These counterrotating gas and stars could
produce an embedded retrograde secondary bar. Friedli (1996) demonstrated
that galaxies having two nested and counterrotating bars are stable and
long-lived systems. He also predicted the peculiar signatures
characterizing their kinematics, which is not yet observed.

\subsection{Gas vs. Gas}

NGC~4826, a nearby and relatively isolated Sab(s) spiral called `Devil-Eye'
or `Black-Eye' contains two nested counterrotating gaseous disks (Braun et
al. 1992). Radio and optical observations (Braun et al. 1994, Rubin 1994a)
revealed an inner disk of about 1 kpc radius containing $\sim10^7$
M$_{\odot}$ in H~{\scriptsize I} and $\sim10^8$ M$_{\odot}$ in H$_2$ and a
counterrotating outer gas disk extending from 1.5 to 11 kpc and containing
$\sim10^8$ M$_{\odot}$ in H~{\scriptsize I}. They are coplanar to the
stellar disk. Stars corotate with the inner gas but beyond the dust lane
less than 5\% of them ($\sim10^8$ M$_{\odot}$) corotate with the outer gas
(Rix et al. 1995). \\ 
The kinematical features of NGC~4826 are interpreted (Rix et al. 1995)
considering an original gas-poor galaxy with prograde gas which slowly
acquires a comparable mass of external retrograde gas. The new
counterrotating gas settles in the outer parts of the stellar disk, leaving
undisturbed the galaxy morphology (Walterbos et al. 1994). Another case we
interpret as constituted by two counterrotating gas components as in
NGC~4826, is the edge-on S0 NGC~5252 (Held et al. 1992) 

The late-type galaxy NGC~253 (Anantharamaiah \& Goss 1996) contains in the
central $\sim150$ pc region three nested structures of ionized gas: a
counterrotating inner disk, an orthogonal rotating ring and an outer ring
rotating in the same sense as the galactic disk. 

In the edge-on S0 NGC~7332 Fisher et al. (1994) detected a second gas
component corotating with the stars in addition to the main counterrotating
gaseous disk. Recently two dimensional high resolution H$\alpha$
spectroscopy (Plana \& Boulesteix 1996) has revealed that the two ionized
gas counterrotating structures are apparently superimposed. They extend to
about 4 kpc and contain $\sim10^{5}$ M$_{\odot}$ in H~{\scriptsize II}. The
accretion of newly supplied counterrotating gas apposed onto an existing
corotating gas disk investigated by Lovelace \& Chou (1996) could be a
possible formation scenario for NGC~7332.

\section{Conclusions}

Counterrotation occurs in a wide variety of forms (gas vs. stars, stars vs.
stars, gas vs. gas). It is present in galaxies of different morphological
types, ranging from ellipticals to early-type disk galaxies. 

The counterrotation of stellar vs. stellar disks is the type of
counterrotation we expect to be prevailing, since it is the end result of
gas vs. stars counterrotation. Therefore its frequency could be very high
if the area described in Fig.~1 is carefully inspected, using the state of
the art analysis of the shape and asymmetries of the line profiles of the
absorption lines. This would indicate that acquisition and merging events
are common phenomena in the history of galaxies.

We are grateful to D. Burstein for useful comments on the manuscript.

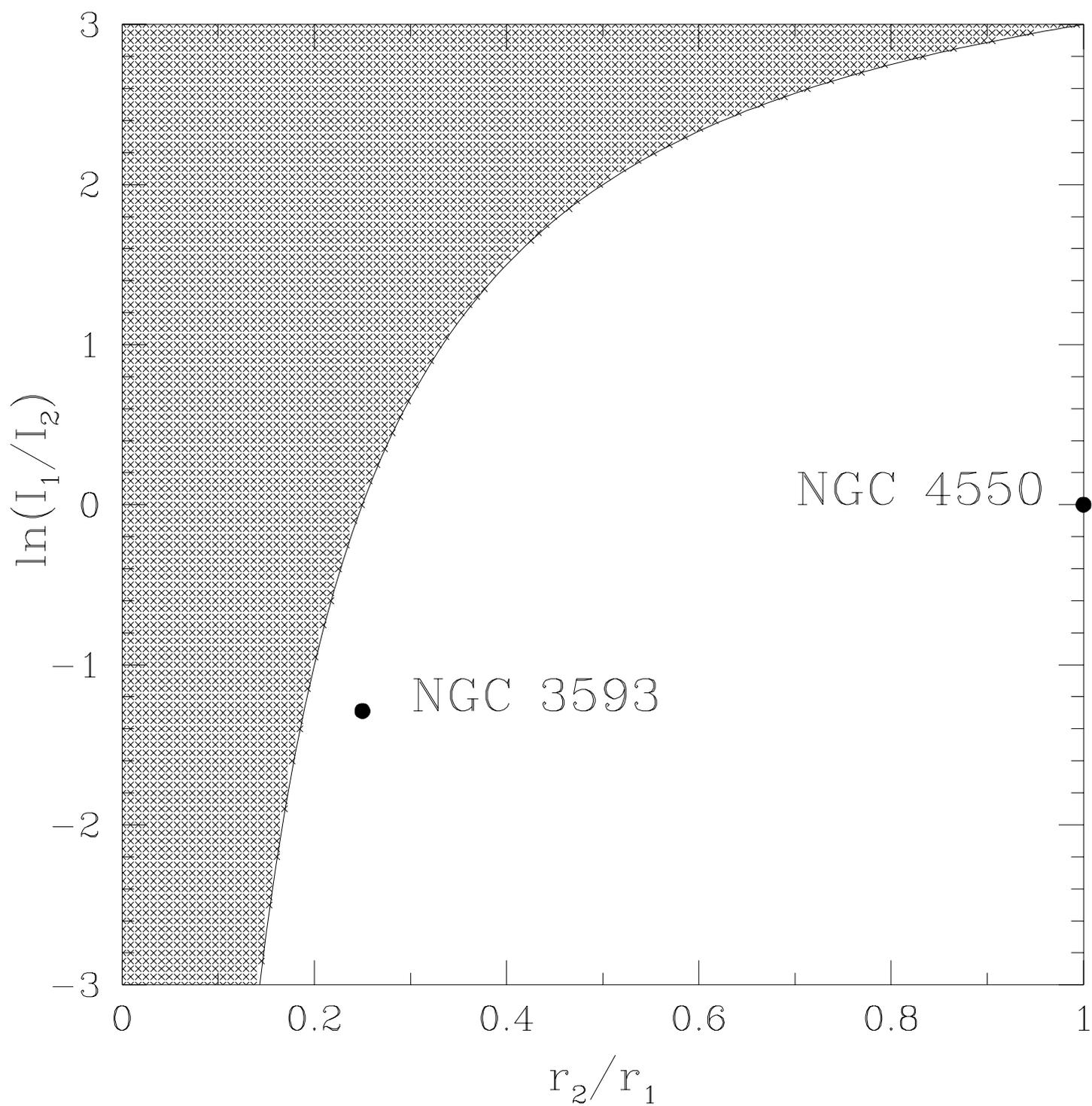
\end{document}